\documentstyle[12pt]{article}
\begin{document}
\thispagestyle{empty}
%
%
\title{Nonlinear  axisymmetric \\
 resistive magnetohydrodynamic equilibria \\ 
  with toroidal flow}

\author{G. N. Throumoulopoulos\thanks{{\it e-mail:} gthroum@cc.uoi.gr}\\
Section of Theoretical Physics,\\ Physics Department, University of
Ioannina,\\
 GR 451 10, Ioannina, Greece}
\date{July  1997}
\maketitle
\begin{abstract}
  The equilibrium of a resistive axisymmetric plasma    
  with purely toroidal flow  surrounded by a conductor
  is investigated within the framework of the nonlinear
  magnetohydrodynamic theory. It is proved that a)
  the poloidal current density  vanishes and b) apart    
  from an idealized case 
  the 
  pressure profile should vanish
  on the plasma boundary.
 For the cases of isothermal magnetic surfaces,
 isentropic magnetic surfaces and magnetic surfaces with constant density,
 the equilibrium states obey to  an elliptic partial
 differential equation for the poloidal magnetic flux function, 
 which is identical in form to the
 corresponding equation governing ideal equilibria. The conductivity,
 which can not be  either uniform or a surface quantity, results 
 however in
 a restriction of the possible classes of equilibrium solutions, e.g. 
 for the cases considered the only possible equilibria with 
 Spitzer conductivity are of cylindrical shape.
\end{abstract}

\noindent
{\bf PACS} numbers: 52.30.Bt, 52.55.-s
\vspace{2cm}

\noindent
To be published in Journal of Plasma Physics
\newpage
\begin{center} 
{\large \bf I.\ \ Introduction}
\end{center}
 
The understanding  of the equilibrium  properties of a magnetically
confined plasma is  one of the  basic  objectives of
fusion research. The major part of the  studies to date has 
 been devoted to static magnetohydrodynamic (MHD) equilibria 
 of an axisymmetric plasma. They are governed by 
a partial elliptic differential  equation for the poloidal
magnetic flux function $\psi$,  which contains 
two surface
quantities [i.e. the pressure $P(\psi)$ and the poloidal current
$I(\psi)$], known as  Grad-Schl\"uter-Shafranov  equation 
(L\"ust and Schl\"uter 1957; Grad 1958; Shafranov 1958).

In the presence of flow the equilibrium becomes more
complicated. 
Although equilibrium studies of flowing plasmas
 began in the mid 1950s (e.g. Grad 1960), since the early 1980s
 there has been an increasing interest in the investigation
 of the equilibrium properties of  plasmas with mass flow 
(Zehrfeld and Green 1972; Maschke and Perrin 1980;
 Morozov and Solov\'ev 1980; Hameiri 1983;
Clemente and Farengo 1984; Semenzato {\it et al} 1984; Kerner
and Tokuda 1987; Throumoulopoulos and Pantis 1989;
Avinash {\it et al} 1992; \.Zelazny {\it et al} 1993; Throumoulopoulos
and Pantis 1996; Throumoulopoulos and Tasso 1997) 
which was motivated by the observation of plasma rotation 
 in many tokamaks heated by neutral beams
 (Suckewer {\it et al} 1979;
Brau {\it et al} 1983; Core {\it et al} 1987; Scott {\it et al} 1989;
Tammen {\it et al} 1994)
With the adoption of a specific equation of state,
e.g. isentropic magnetic surfaces (Morozov and Solov\'ev 1980),
 the axisymmetric  
ideal MHD equilibrium states obey to a partial differential
equation for  $\psi$, 
containing five surface quantities,  
in conjunction with a nonlinear algebraic
Bernouli equation. Unlike the case in static equilibria, the  
above mentioned differential  equation
is not always elliptic; there are three critical values 
of the poloidal flow 
at which the type of this  equation changes, i.e. 
it becomes alternatively elliptic and hyperbolic. To solve the 
equilibrium problem in the two elliptic regions, several computer
codes have been developed  (Semenzato {\it et al} 1984; Kerner and Tokuda
1987; \.Zelazny {\it et al} 1993). 
There is experimental evidence, however, that either the
flow is purely toroidal
 (Suckewer {\it et al} 1979;
 Core {\it et al} 1987; Scott {\it et al} 1989;
Tammen {\it et al} 1994)
 or  its poloidal component is
one order of magnitude lower than the toroidal one (Brau {\it et al} 1983). 
It was also found that the poloidal flow is
efficiently  damped by magnetic pumping  (Hassam and Kulsrud 1978).
In addition,  toroidal rotation has been suggested 
as a possible means of creating a magnetic field in the 
sun and stars (Plumpton and Ferraro 1955)  and of stabilizing   
the tilting (Hayashi and Sato 1987),  ballooning  (Miller {\it et al} 1995)
 and
drift-like (Sen and Rusbridge 1995) modes. 
Therefore toroidal flow
may be considered a dominant effect.
For a purely toroidal velocity component
the partial differential equation becomes elliptic and can be 
solved analytically (Maschke and Perrin 1980; Clemente and Farengo 1984;
Throumoulopoulos and Pantis 1989). 

The investigation of the general MHD equilibrium, involving  plasma flow,  
finite conductivity and viscosity is a very difficult problem, particularly
in light of the uncertainties in the viscous stress tensor. For 
example, for the case in which the mean free path is greater than 
a gyroradius, as it is in the case of a tokamak, 
the ``ion parallel viscosity" (Braginskii 1965; Balescu 1988)
takes an extraordinary  
large value, so large as to make the applicability of either
version of the MHD viscosity stress tensor to tokamak dynamics
questionable. It may be noted, however, that the ion parallel viscosity 
first increases strongly with the temperature until the mean free
path reaches the order of magnitude of the machine, then decreases
for larger values of the mean free path. In particular, in the high
viscosity regime
resistive equilibria with flows originated from paired toroidal 
vortices, which  are very likely created in a toroidal magnetofluid, 
were investigated recently (Montgomery Bates and Li 1997a);
the existence
of such vortices is however conjectured to characterize magnetofluids  
beyond the high-viscosity limit.
Thus, the form and the  magnitude of 
the appropriate viscous stress tensor to be used in tokamak MHD has been
the subject of considerable discussion, and at present these
discussions show no signs of converging.

For an inviscid plasma 
Pfirsch and Schl\"uter (Pfirsch and Schl\"uter 1962; see also Wesson 1987)   
showed long time ago that in the collisional regime  
the toroidal curvature gives  rise to an enhanced diffusion, which is 
related to the  conductivity parallel to the magnetic field. Since then 
there have been attempts to examine the effects of finite
conductivity (e.g. Grad and Hogan 1970), but these attempts
appear  not having   led  to a satisfactory resolution;
too many possibilities are raised
to be able to deal with any one of them conclusively.
In the above mentioned studies
the inertial-force flow term 
$\rho({\bf v}\cdot\nabla){\bf v}$  is neglected in the equation of 
momentum conservation.
For ion flow velocities of the order of $10^7$ cm/sec,
 which have
been observed
in neutral-beam-heating experiments (Suckewer {\it et al} 1979;
Brau 1983 {\it et al}; Tammen {\it et al} 1994), however,
the term
$\rho({\bf v}\cdot\nabla){\bf v}$ 
can not be considered negligible. Therefore, it is
worthwhile to investigate the nonlinear 
resistive equilibrium.   
 Such an investigation   is  
also  encouraged  by a proof (Tasso 1979) according
to which 
a conductivity profile
remaining constant on a magnetic surface is not  
compatible with the  Grad-Schl\"uter-Shafranov equation. 
The non existence
of static axisymmetric resistive equilibria with a uniform conductivity was
also  suggested  recently (Montgomery and Shan 1994; Bates and Lewis
1996; Montgomery {\it et al} 1997b).   

In the present paper we study 
the  MHD equilibrium 
of an axisymmetric plasma
with scalar conductivity and purely toroidal  flow 
surrounded 
by a conductor,
including the term  $\rho({\bf v}\cdot\nabla){\bf v}$
in the momentum conservation equation.
It is found that the presence of
 flow does not remove the above
mentioned incompatibilities  associated with the conductivity. 
Also, the conductivity has an impact  
on the direction of the current density,
on the boundary conditions  and 
restricts the possible  classes  of  equilibrium solutions.

The paper  is organized as follows.
For the system under consideration equilibrium equations 
independent of the plasma equation of state are derived  in Sec. II. 
The   cases of equilibria with (a) isothermal (b) isentropic 
and (c) constant density magnetic surfaces  are examined in Sec. III.
Sec. IV summarizes our conclusions.

\begin{center}
{\large\bf II.\ \ Equilibrium equations }
\end{center}

The MHD equilibrium states of an inviscid plasma
with scalar conductivity
are governed by the following set of
equations, written in standard notations and convenient units:
\begin{equation}
{\bf\nabla} \cdot (\rho {\bf v}) = 0 
					    \label{1}
\end{equation}
\begin{equation}
\rho ({\bf v} \cdot {\bf\nabla})  {\bf v} = {\bf j}
\times {\bf B} - {\bf\nabla} P 
					    \label{2}
\end{equation}
\begin{equation}
{\bf\nabla} \times  {\bf E} = 0 
					    \label{3}
\end{equation}
\begin{equation}
{\bf\nabla}\times {\bf B} = {\bf j }
					    \label{4}
\end{equation}
\begin{equation}
{\bf\nabla} \cdot {\bf B} = 0 
					    \label{5}
\end{equation}
\begin{equation}
{\bf E} +{\bf v} \times {\bf B} = \frac{\bf j}{\sigma}.
					    \label{6}
\end{equation}
It is pointed out that, unlike to the usual procedure  
followed in equilibrium studies  with flow
(Zehrfeld and Green 1972; Morozov and Solov\'ev 1980; Hameiri 1983; 
Semenzato Gruber and Zehrfeld 1984;
Kerner and Tokuda 1987; \.Zelazny {\it et al} 1993) 
in the present work
an equation of state is not included
in the above set of  equations from the outset  and therefore 
the equation of state independent Eqs. (\ref{14}) and (\ref{15})
below are first derived.  This alternative procedure is convenient
because the equilibrium problem is further reduced  then
for specific   cases associated  with several  equations of state.

The system under consideration 
is a toroidal axisymmetric magnetically confined   plasma, 
which is surrounded by a conductor (Fig. 1). Cylindrical coordinates
$R, \phi, z$ are employed with unit basis vectors ${\bf e}_R, 
{\bf e}_\phi,  {\bf  e}_z$, where  ${\bf  e}_z$
is parallel to the axis of symmetry. The position
of the surface of the conductor   is specified by some 
boundary curve in the ($R$, $z$) plane. 
The equilibrium quantities do not depend on the azimuthal 
coordinate $\phi$.
Also, it is assumed that  the plasma elements flow solely
along ${\bf  e}_\phi$. Eq. (\ref{1})  is then identically
satisfied.
 
For the above mentioned  system we first prove that the equilibrium poloidal
current density  must vanish. Inserting the  expression
\begin{equation}
{\bf B} =B_\phi {\bf e}_\phi + {\bf B}_{\mbox{pol}} =  
	 B_\phi {\bf e}_\phi +\frac{1}{R}{\bf\nabla} \psi 
	 \times {\bf e}_\phi
				       \label{7}
\end{equation}
for the magnetic field 
into Ampere's law (\ref{4})                                               
yields  for the current
density:
\begin{equation}
{\bf j} = -\frac{1}{R}\Delta^\star\psi {\bf e}_\phi 
	  + \frac{1}{R}\nabla (RB_\phi) 
	  \times {\bf e}_\phi. 
					 \label{8}
\end{equation}
Here, $\psi$ is the  poloidal magnetic flux function
which labels the magnetic surfaces, and $\Delta^\star $ is the elliptic 
operator defined by $\Delta\psi = R^2\nabla\cdot\left(\nabla\psi/R^2\right)$.
With the aid of Eqs. (\ref{7}) and (\ref{8}), the component of the
force balance equation (\ref{2}) along ${\bf e}_\phi$
is written in the form 
\begin{equation}
{\bf e}_\phi \cdot \nabla(RB_\phi)\times\nabla\psi=0.
					      \label{9}
\end{equation}
Eq. (\ref{9}) implies that  
$RB_\phi$ is a surface quantity: 
\begin{equation}
RB_\phi\equiv I(\psi).
					       \label{9a}
\end{equation}
Integration of  Ohm's law (\ref{6}) along a 
contour $c$ defined by the cut of an
arbitrary closed magnetic surface with the poloidal plane (Fig. 1)
leads to the equation
\begin{equation}
\int_c {\bf E}\cdot d{\bf l} + \int_c ({\bf v}\times {\bf B})                                         
\cdot d {\bf l} = \int_c\frac{1}{\sigma}({\bf j}\cdot d {\bf l}).
					       \label{10}
\end{equation}
Since in equilibrium it holds that $\partial {\bf B}/\partial t = 0$, 
the first
integral vanishes by Stokes's theorem. Also, the second integral 
vanishes  because  ${\bf v}$   is purely toroidal. 
Consequently, the term on the right-hand side
must vanish,  implying that
 ${\bf j}_{\mbox{pol}}\equiv 0$ and therefore $I=$ constant. 

The $\phi$-component of the electric field associated with 
the current density can be
obtained   from Faraday's law  
(\ref{3}):  
\begin{equation}
E_\phi=\frac{j_\phi}{\sigma}=E_0\frac{R_0}{R},
					    \label{11}
\end{equation}
where $E_0$ is the value of the electric filed  
in the cylindrical surface $R=R_0$.
An electric field of this kind can be created by a linear change in
time of the magnetic flux  confined to an axisymmetric
very long iron core, which is directed along ${\bf e}_z$
and is located within the hole of the torus.  This
can be considered as a highly simplified representation  
of a tokamak. 
 
Eq. (\ref{11}) has an impact on the boundary conditions,
i.e. the component of ${\bf E}$
tangential to the plasma-conductor interface does not vanish. 
Therefore, apart from a special idealized situation  
in which the surface of  a perfectly conducting
container is coated by a very thin layer of insulating dielectric
(Bates and Lewis 1996; Montgomery et al 1997b) 
 the container can not be considered  perfectly
conducting.
Accordingly,  Ohm's law with  finite conductivity applied in the
vicinity of the plasma-conductor interface does not permit
the existence of a surface layer of current.
It is now assumed that the position of the conductor is such that
 its surface 
coincides with the outermost of the closed magnetic surfaces. 
Thus, 
 the condition
 ${\bf B}\cdot {\bf n}=0$, where {\bf n} is an outward unit vector
 normal to the plasma surface (Fig. 1),  holds in  the plasma-conductor
 interface.  
 It is noticed
 that this  is possible  only in equilibrium, 
 because in the context of resistive MHD time dependent
 equations, the magnetic flux is not conserved. Integration
 of the momentum conservation  equation 
 across the plasma-conductor interface and
 projection of the resulting equation parallel 
 to {\bf n} yields (with {\bf B} being continuous
 in the plasma-conductor interface)
 \begin{equation}
P + \rho({\bf n}\cdot{\bf v})^2 = 0. 
					    \label{11a}
\end{equation}
Since all the terms of Eq. (\ref{11a}) are non-negative
and the flow is purely toroidal, it follows that the pressure
$P$ must vanish at the plasma boundary.

To  reduce the equilibrium equations (\ref{1})-(\ref{6}), a second
surface quantity is identified by projecting the equation
\begin{equation}
\nabla\times \left({\bf v}\times {\bf B} - \frac{1}{\sigma}j_\phi
{\bf e}_\phi\right)=0
					      \label{12}
\end{equation}
[following from a combination of Eqs. (\ref{3}) and (\ref{6})]   
along ${\bf e}_\phi$. This yields
\begin{equation}
{\bf e}_\phi\cdot \nabla\left(\frac{v_\phi}{R}\right)
\times\nabla\psi = 0.
					      \label{13}
\end{equation}
Equation (\ref{13}) implies that the angular frequency  of the flow
shares the same surfaces with $\psi$: 
\begin{equation}
\omega\equiv\frac{v_\phi}{R}=\omega({\psi}).
					      \label{13a}
\end{equation}
With the aid of equations  (\ref{7}), (\ref{8}) with $I=$ const., 
(\ref{9a}) and (\ref{13a}),
the components of Eq. (\ref{2}) along ${\bf B}_{\mbox{pol}}$ and perpendicular
to a magnetic surface are put in the forms, respectively,
\begin{equation}
{\bf B}_{\mbox{pol}}\cdot\left[\frac{\nabla P}{\rho} -
\nabla\left(\frac{\omega^2R^2}{2}\right)\right] = 0
					      \label{14}
\end{equation}
and
\begin{equation}
\Delta^\star \psi +R^2\frac{\nabla \psi}{\left|\nabla \psi\right|^2}                                              
\cdot \left[\nabla P - \rho\omega^2 
\nabla\left(\frac{R^2}{2}\right)\right] = 0.
					      \label{15}
\end{equation}
It is noted here that Eqs. (\ref{9}), (\ref{14}) and (\ref{15})
are orthogonal components of Eq. (\ref{2}) and therefore 
they are independent. 

Summarizing, the resistive MHD equilibrium of an axisymmetric
toroidal plasma with purely toroidal flow is governed by the
set of Eqs. (\ref{14}), {\ref{15}) and
\begin{equation}
\Delta^\star \psi +R_0E_0\sigma=0,
					      \label{15a}
\end{equation}
following from Eqs. (\ref{8}) and (\ref{11}).  
Owing to the purely toroidal direction of the flow, which
leads to ${\bf j}_{\mbox{pol}}=0$ and $\omega=\omega(\psi)$,
Eqs. (\ref{14}) and (\ref{15}) {\em do not contain the 
conductivity} and are identical in form to the corresponding
equations governing ideal equilibria. 
Therefore, on the one side,
several properties  of the ideal equilibria, e.g. 
the Shafranov shift of the magnetic surfaces and the detachment
of the isobaric surfaces from the magnetic surfaces (see the 
discussion following Eq. (\ref{17}) below) remain valid.
On the other side, 
Eq. (\ref{15a})
can be employed as an algebraic equation for the conductivity
i.e. the solutions of the  ``ideal MHD"  equations (\ref{14})
and ({\ref{15}) can determine the spatial dependence of
$\sigma$   
by means of the formula
\begin{equation}
\sigma=-\frac{1}{R_0 E_0}\Delta \psi^\star.
					      \label{16}
\end{equation}
Eq. (\ref{16}) implies that the conductivity can
not be  uniform, as the case in static equilibria
(Tasso 1979; Montgomery and Shan 1994; Bates and Lewis 1996;
Montgomery {\it et al} 1997b).

It may be noted here that the 
 equilibrium problem
can be treated in a different way, suggested for the case
of static resistive equilibria (Bates and Lewis 1996),  
 i.e. Eq. (\ref{15a}) 
is employed as an elliptic partial differential equation
for $\psi$, which is solved after the spatial dependence of $\sigma$ 
is assigned. 
To our opinion, this method has two drawbacks: a) A physically
unjustifiable assignment of conductivity is required from the 
outset. b) The conductivity profile must be compatible with
Eq. (\ref{15}), a requirement which (at least in the presence of flow) makes
the assignment difficult. This becomes more clear in next section
[see Eqs. (\ref{18}), (\ref{22b}) and (\ref{25})].

\begin{center}
{\large\bf III Magnetic surfaces with  special properties}
\end{center}

In order to further reduce  the equilibrium
equations (\ref{14}) and (\ref{15}) and find out
their influence  on the conductivity profile,  the starting set
of equations   (\ref{1})-(\ref{6})
 must be supplemented by 
an  equation of state, e.g. $P=P(\rho, T)$,  
along with
an equation determining the transport
of internal energy such as
\begin{equation}
C_v\left(\frac{\partial}{\partial t}
	 + {\bf v}\cdot\nabla\right) T
	 + P\nabla\cdot {\bf v} + \kappa\nabla^2 T =0,
					   \label{16a}
\end{equation}
the energy transport equation for a perfect gas, where $C_v$ is
the volume specific heat, $\kappa$ the thermal conductivity
and $\frac{\textstyle\partial T}{\textstyle\partial t}=0$.
Such a rigorous treatment, however,  makes the equilibrium problem 
very cumbersome. Alternatively, since for 
fusion plasmas the thermal conduction along $\bf B$ is expected
to  be fast in relation to  the heat transport perpendicular
to a magnetic surface,  equilibria
with  isothermal magnetic surfaces is a reasonable approximation
 (e.g. Mashke and Perrin 1980; Clemente and Farengo 1984;
Throumoulopoulos and Pantis 1989).
In particular, the even simpler case of 
isothermal resistive equilibria ($T=\mbox{constant}$)
 has also been considered (Grad and Hogan 1970).
Equilibria with isothermal
magnetic surfaces are examined as follows.

\begin{center}
{\large\em  Equilibria with isothermal magnetic surfaces}
\end{center}

 In addition to $T=T(\psi)$ the plasma is assumed to obey the perfect gas
law $P = \lambda \rho T$.
  For this kind of equilibria, Eq. (\ref{14}) can be integrated
yielding an expression for the pressure:
\begin{equation} 
P = P_T(\psi)\exp\left(\frac{R^2\omega^2}{2\lambda T}\right).
					    \label{17}    
\end{equation}
We note here that, unlike in static 
equilibria, in the presence of flow  magnetic surfaces do not coincide with
isobaric surfaces  because Eq. (\ref{2})
implies that ${\bf B} \cdot {\bf \nabla} P$ in
general differs from zero.
In this respect, the term $P_T(\psi)$ is 
the static part of the pressure ($\omega=0$).
Inserting Eq. (\ref{17}) into Eq. (\ref{15}), the latter
takes the form 
\begin{equation}
\Delta^\star\psi 
+ R^2\left[P_T^\prime + P_T\frac{R^2}{2}
\left(\frac{\omega^2}{\lambda T}
\right)^\prime\right]\exp\left(\frac{R^2\omega^2}{2\lambda T}\right) = 0,
						  \label{18}
\end{equation}
where the prime  denotes differentiation with respect to $\psi$.
This   
is the  equilibrium equation for a resistive plasma with 
purely toroidal flow and isothermal magnetic surfaces. 
It contains the three arbitrary
surface quantities $T(\psi)$, $\omega(\psi)$ and $P_T(\psi)$, 
which must be found from other physical considerations.
For $\omega\equiv 0$  it reduces to
the Grad-Schl\"uter-Shafranov equation for vanishing
poloidal current.  For ideal equilibria,  
Eq. (\ref{18})    was obtained by Maschke and
Perrin (1980).  For special  choices of the surface quantities,
which must lead to physically reasonable profiles for the equilibrium
quantities,  Eq. (\ref{18}) can be linearized and solved analytically.
In particular, with the ansatz
$\omega^2/(2\lambda T) \equiv\omega_T=$ const.
it takes the simpler form
\begin{equation}
\Delta^\star\psi + R^2P_T(\psi)^\prime\exp\left(R^2\omega_T
						\right).
						\label{19}
\end{equation}
 Solutions of Eq.  (\ref{19}) have been obtained for $P_T\propto\psi$
 (Maschke and Perrin 1980)  and for $P_T\propto\psi^2$
 (Clemente  Farengo 1984; Throumoulopoulos and Pantis 1989)
 and the detachment of 
 the isobaric surfaces from the magnetic surfaces was evaluated. 
 
 Eqs.  (\ref{16}) and (\ref{18}) yield
\begin{equation}
E_0R_0\sigma= R^2\left[P_T^\prime + P_T\frac{R^2}{2}
\left(\frac{\omega^2}{\lambda T}
\right)^\prime\right]\exp\left(\frac{R^2\omega^2}{2\lambda T}\right). 
							  \label{19a}
\end{equation}
As the case in static equilibria (Tasso 1970; Montgomery and Shan 1994)
toroidal equilibria with the  conductivity remaining constant
in a magnetic surface are not possible. Indeed, if $\sigma=\sigma(\psi)$
the left-hand side of Eq. (\ref{19a})  becomes  a surface quantity, 
while the right-hand side
(in addition to surface quantities)  involves an explicit dependence
of $R$. Considering the variables $R$ and $\psi$, instead of  $R$ and $z$,  
as independent [then $z=z(R,\psi$)], Eq. (\ref{19a}), after expanding
  the exponential term  in  a Taylor series,
  can be  written
 in the form $\sum_{n=0}^{\infty} a_n(\psi) R^n =0 $.  The latter 
requires that
the relation $a_n(\psi)\equiv 0$ must
be satisfied for all  $n$. This is impossible because
$a_0=E_0R_0\sigma(\psi)$ and therefore it follows that $\psi=\psi(R)$. 
Thus,
the only possible $\sigma=\sigma(\psi)$ equilibria are of cylindrical shape.
It may be noted here that the non existence
of another class of resistive toroidal MHD equilibria,  
i.e. axisymmetric incompressible $\beta_p=1$
equilibria  with  purely poloidal 
flow, was proved by Tasso (1970).
 
To find out whether the above incompatibilities   
associated with the  conductivity profile 
arise from the choice $T=T(\psi)$, one 
alternatively can consider 
the  cases
of  isentropic processes and incompressible flows,
 which has been the subject of extensive ideal MHD studies and are,
 respectively,  associated with isentropic magnetic surfaces  
(Maschke and Perrin 1980; Morozov and Solov\'ev 1980; Hameiri 1983;
 Semenzato {\it et al} 1984}; Kerner
and Tokuda 1987; Throumoulopoulos and Pantis 1989; \.Zelazny {\it et al} 1993)
and constant density magnetic surfaces 
(Avinash {\it et al} 1992;  Throumoulopoulos
and Pantis 1996; Throumoulopoulos and Tasso 1997).

\begin{center}
{\large\em Equilibria with isentropic magnetic surfaces}
\end{center}

We consider a plasma with large but  
finite conductivity such that for  times  short compared with
the diffusion time scale, the dissipative   term
$\approx j^2/\sigma$ can be neglected.  
This permits one to  assume  
that the magnetic surfaces are isentropic, i.e $S=S(\psi)$
where $S$  is the plasma specific entropy. It may be noted that
 $S=S(\psi)$  can be proved 
for the stationary states of an ideal plasma with  arbitrary flows
on the basis of the entropy conservation 
(Morozov and Solov\'ev 1980; Hameiri 1983). 
 This case was also studied for ideal equilibria 
 with purely toroidal flow (Maschke and Perrin 1980; Throumoulopoulos
and Pantis 1989).
 In addition, the plasma is assumed  
to being a perfect gas whose
internal energy density $E$ is simply proportional to the temperature.
Then, the  equations  for the  thermodynamic potentials lead to  
(Maschke and Perrin 1980)
\begin{equation}
P = A(S)\rho^\gamma
						      \label{21}
\end{equation}
and
\begin{equation}
E =\frac{A(S)}{\gamma -1}\rho^{\gamma -1} = \frac{H}{\gamma}.
						      \label{22}
\end{equation}
Here, $A=A(S)$ is  an arbitrary function of the specific entropy 
$S\equiv S(\psi)$,
$ H=E + P/\rho$ is the specific enthalpy 
and $\gamma$ is the ratio of specific heats.
For simplicity and without loss of generality we choose the function $A$
to be identical with $S$. Consequently,
integration of Eq. (\ref{14})  yields
\begin{equation}
H -\frac{R^2\omega^2}{2} \equiv\theta(\psi),
						 \label{22a}
\end{equation}
where $\theta$ is a surface quantity.
Eq. (\ref{15}) reduces then to
\begin{eqnarray}
\lefteqn{\Delta^\star\psi + 
R^2\left(1 +\frac{R^2\omega^2}{2\theta}\right)^{\eta-1} } & & \nonumber \\[3mm]
& \times& \left\{\left[\left(\frac{\theta}{\eta}\right) S^{1-\eta}\right]^\prime        
+ R^2\left[\left\{\left(\frac{\theta}{\eta}\right)^\eta\right\}^\prime
  S^{1-\eta}\frac{\omega\omega^\prime}{\theta^\prime}
  + \left(\frac{\theta}{\eta}\right)^\eta (S^{1-\eta})^\prime
  \frac{\omega^2}{2\theta}\right]\right\}=0,   
						      \label{22b}
\end{eqnarray}
where $\eta=\gamma/(\gamma-1)$.
This is the resistive equilibrium equation for isentropic
magnetic surfaces, which is the same as the ideal equation 
 (Maschke and Perrin 1980; Throumoulopoulos and Pantis 1989)
 with $j_{\mbox{pol}}= 0$.
 The three surface quantities 
are now $\omega(\psi)$,  $S(\psi)$ and $\theta(\psi)$. With the ansatz
$\omega^2/(2\theta) =\omega_s=$ const., Eq. (\ref{22b}) is  
put in the simpler form
\begin{equation}
\Delta^\star\psi + R^2\left(1 + \omega_s R^2\right)^\eta (P_s)^\prime 
= 0,
						      \label{23}
\end{equation}
where the function  
$$
P_s(\psi) \equiv \left(\frac{\theta}{\eta}\right)^\eta S^{1-\eta} 
$$
stands for the static equilibrium
 pressure.
Solutions of Eq. (\ref{23}) are available in the literature
for $P_s\propto\psi$ (Maschke and Perrin 1980) and for $P_s\propto\psi^2$
 (Throumoulopoulos and Pantis 1989). 

Eqs. (\ref{16}) and (\ref{22b}) imply  that 
toroidal isentropic equilibria
are not compatible with $\sigma=\sigma(\psi)$.
In addition, 
equilibria with Spitzer conductivity 
$\sigma=a T^{3/2}$
 satisfy the equation
\begin{eqnarray}
\lefteqn{R_0E_0a\left[\frac{1}{\gamma S}\left(\theta + \frac{R^2\omega^2}{2}
  \right)\right]^{3/2} = 
  R^2\left(1 +\frac{R^2\omega^2}{2\theta}\right)^{\eta-1} } & & \nonumber \\
  &\times & \left\{\left[\left(\frac{\theta}{\eta}\right) 
  S^{1-\eta}\right]^\prime        
+ R^2\left[\left\{\left(\frac{\theta}{\eta}\right)^\eta\right\}^\prime
  S^{1-\eta}\frac{\omega\omega^\prime}{\theta^\prime}
  + \left(\frac{\theta}{\eta}\right)^\eta (S^{1-\eta})^\prime
  \frac{\omega^2}{2\theta}\right]\right\}=0,   
							 \label{24}
\end{eqnarray}
following from Eqs. (\ref{16}) and ({\ref{21}-\ref{22b}).
Eq. (\ref{24}) contains, in addition to 
 surface quantities, the coordinate  $R$ explicitly. 
Therefore, as in the  case of isothermal magnetic surfaces and
$\sigma=\sigma(\psi)$, equilibria with isentropic magnetic 
surfaces and Spitzer conductivity must have cylindrical shape.

\begin{center}
{\large \em Equilibria with magnetic surfaces of constant density} 
\end{center}

Equilibria with incompressible flows 
are usually associated with constant density magnetic surfaces
(Avinash {\it et al} 1992;
Throumoulopoulos and Pantis 1996; Throumoulopoulos and Tasso 1997).
For 
axisymmetric equilibria with purely toroidal flow, Eqs.
 (\ref{1}) and  $\nabla \cdot {\bf v}=0$
are satisfied identically and provide no information about
$\rho$. Nevertheless, $\rho=\rho(\psi)$   
 is a physically possible  equation of state. 
For this case, integration of  Eq. ({\ref{14}) leads to 
\begin{equation}
P=P_\rho(\psi) + \frac{\rho R^2 \omega^2}{2},
					     \label{24a}
\end{equation}
where $P_\rho(\psi)$ is the static equilibrium pressure.
Consequently, Eq. (\ref{15}) reduces to
\begin{equation}
\Delta^\star\psi + R^2\left[P_\rho^\prime 
+\frac{R^2}{2}\left(\rho\omega^2\right)^\prime\right] = 0.
					       \label{25}
\end{equation}
As in the cases of isothermal and isentropic magnetic
surfaces,
Eqs. (\ref{16}) and (\ref{25}) imply that in toroidal equilibria 
the conductivity can not be
a surface quantity. Also,  equilibria with Spitzer conductivity
satisfy the relation
\begin{equation}
E_0R_0a\left(\frac{P_\rho}{\lambda\rho} + \frac{\omega^2 R^2}{2\lambda}
\right)^{3/2}
=R^2\left[P_\rho^\prime + 
\frac{R^2}{2}\left(\rho\omega^2\right)^\prime\right], 
					       \label{26}
\end{equation}
which implies that magnetic surfaces must have cylindrical shape.

Concluding this subsection, we 
derive two classes of exact solutions of Eq. (\ref{25}).
With the use of the ansatz
\begin{equation}
\rho\omega^2=\mbox{const},                                               
					       \label{27}
\end{equation}
Eq. (\ref{25}) reduces to
\begin{equation}
\Delta^\star\psi +R^2 P_\rho^\prime=0,                                             
					       \label{27a}
\end{equation}
which is identical in form to the 
the Grad-Schl\"uter-Shafranov 
equation for vanishing poloidal current. 
A simple solution of Eq. (\ref{27a}) can be derived
for $P_\rho\propto\psi$ (Shafranov 1966):
\begin{equation}
\psi =\psi_c\frac{R^2}{R_c^4}\left(2R_c^2 -R^2 -4d^2 z^2\right).                                           
					       \label{27b}
\end{equation}                                                                  
Here, $\psi_c$ is the value of the flux function at the position 
 of the magnetic axis ($z=0$, $R=R_c$)
 and $d^2$ is a parameter which determines the shape
of the flux surfaces,
e.g. for $d^2=1$ Eq. (\ref{27b}) describes a compact torus contained 
within a spherical conductor.  
For $P_\rho\propto\psi^2$, solutions of Eq. (\ref{27a}) 
separable
in the coordinates $R$ and $z$ can be expressed
in terms of Coulomb wave functions  and describe
equilibria of either  a tokamak  with rectangular cross section 
(Hernegger 1972; Maschke 1973)  or a compact torus contained
 within a rectangular
conductor (Berk {\it at al} 1981).
These solutions  can yield  a peaked current density profile vanishing 
on the boundary. 
Owing to the presence of the flow term in Eq. (\ref{24a}) however
the pressure does not vanish on the boundary, as it should 
for resistive equilibria. Thus, 
solutions of this kind are not valid in the vicinity of the boundary.
The drawbacks of the solutions presented  above in this paragraph,
arise from the choices (\ref{27}) and $P_\rho\propto\psi $. They can be 
eliminated 
if, alternatively,  the   ansatz
$
\rho\omega^2=k\psi^2$ and $P_\rho=\mu\psi^2/2 $                                                
is employed, where  $k$ and $\mu$ are constants. 
Then, Eq. (\ref{25})  reduces to
\begin{equation}
\Delta^\star\psi + R^2\left[\mu + kR^2\right]\psi= 0.
					       \label{27c}
\end{equation}
The last equation
can not be solved analytically and numerical solutions will
not be derived here. 

\begin{center}
{\large\bf IV.\ \ Conclusions}
\end{center}

The  equilibrium of an axisymmetric plasma 
with  scalar conductivity and purely toroidal flow 
surrounded by a conductor was investigated within the 
framework of the nonlinear MHD theory. It was found that a) the poloidal
current density ${\bf j}_{\mbox{pol}}$ must vanish and 
b) apart from an idealized case,
perfect-conducting-boundary conditions are not relevant.  
The equilibrium states are governed
by a set of two differential 
equations [Eqs. (\ref{14}) and (\ref{15})], which 
do not contain the conductivity and are identical in form to
the corresponding equations for ideal stationary equilibria. 
The conductivity, however,  restricts the allowed  equilibria
to those that satisfy ${\bf j}_{\mbox{pol}}=0$ 
and
have pressure profiles vanishing on the boundary. 
With the assumption of a specific equation
of state, the above mentioned set of equations reduces to a single
elliptic partial differential equation for the poloidal magnetic flux
function which contains three surface quantities [e.g. Eqs. (\ref{18}), 
(\ref{22b}), (\ref{25})]. 
The cases
of isothermal magnetic surfaces, isentropic magnetic surfaces and
magnetic surfaces of constant density were examined and, 
for specific choices of the surface quantities involved in each
case, equilibrium solutions were presented and discussed.

As the case in
static equilibria,
the conductivity can not be uniform.  
In addition,  for all three cases of equation
of state examined, the conductivity  
can not be a surface quantity;  
the only possible equilibria with conductivity remaining 
constant in a magnetic surface  have cylindrical 
shape.
Also, for isentropic magnetic surfaces
and magnetic surfaces of constant density,
it was shown 
the non existence of toroidal equilibria with Spitzer conductivity.

It is  interesting to investigate whether these 
unusual properties of conductivity  persist when 
 additional physical input, e.g. a general flow with  non vanishing
poloidal and toroidal components is included. Of course,   pursuing
such kind of investigations, one gets a step closer 
to the cumbersome problem of MHD equilibrium 
with flow, finite conductivity and viscosity. 
The results of the few  studies which have been conducted  to date              
indicate, however, that 
this  objective  may
be of crucial importance for fusion plasmas.
\begin{center}
 {\large\bf Acknowledgments}
\end{center}

The author would like to thank Dr. H. Tasso for
a  valuable discussion,  for providing his report
(Tasso 1979) 
and a critical reading of the manuscript. The referees'  comments 
which were helpful in putting the manuscript in its final form
are acknowledged. Part of the  work was conducted during 
a visit of the author to  Max-Planck Institute f\"ur Plasmaphysik, 
Garching, Germany. The hospitality provided at  said institute 
is appreciated.

Support by EURATOM (Mobility Contract No
131-83-7 FUSC) is  acknowledged.
\begin{center}
 {\large\bf References}
\end{center}
AVINASH, K \&  BHATTACHARYYA, S. N. \& GREEN, B. J. 1992
	       {\it Plasma Phys. Control. Fusion} {\bf 34}, 465.\\ 
BALESCU, R. 1988 {\it Transport Processes in Plasmas}, 
	     (Amsterdam, North-Holland).\\
BATES, J. W. \& LEWIS, H. R. 1996  {\it Phys. Plasmas} {\bf 3}, 2395. \\
BERK, H. L. \&  HAMMER, J. H. \& WEITZNER, H 1981 
	       {\it Phys. Fluids}  {\bf 24},
		 1758. \\                   
BRAGINSKII, S. I. 1965   {\it Reviews of Plasma Physics},
	       edited by M. A. Leontovich (New York, 
	      Consultants' Bureau) Vol. 1,  p. 205.\\ 
BRAU, K. \& BITTER, M. \& GOLDSTON, R. J. \& MANOS, D. \& 
     McGUIRE, K. \& SUCKEWER, S. 1983 {\it Nucl. Fusion} {\bf 23}, 1643.\\
CLEMENTE, R. A. \& FARENGO, R.  1984   {\it Phys. Fluids}
		{\bf 27}, 776.\\ 
CORE, W. G. F. \& P. VAN BELL, P. \&  Sadler G. 1987   
		    {\it  Proc. of the 14st European Conf.
                   Controlled Fusion and
		  Plasma Physics (Madrid, 1987)}, 
		  (Geneva: European Physical Society)
		  vol 11D,  p 49.  \\ 
GRAD, H., 1960    {\it Rev. Mod. Phys.} {\bf 32}, 830. \\  
GRAD, H.  \&  RUBIN, H.  1958 
	      {\it Proceedings of the Second United Nations Conference on  
	      the Peaceful Uses of Atomic Energy, Geneva, 1958}, edited by
	      United Nations (United Nations Publication, Geneva), 
	      Vol. 31, p. 190.  \\
GRAD, H. \& HOGAN, J. 1970  {\it Phys. Rev. Lett.} {\bf 24}, 1337.\\
HAMEIRI, E.  {\it Phys. Fluids} 1983 {\bf 26}, 230.\\ 
HASSAM, A. B. \&   KULSRUD, R. M. 1978 {\it Phys. Fluids} {\bf 21}, 2271.\\ 
HAYASHI, T. \&  SATO, T. 1987 {\it J. Phys. Soc. Jpn.} {\bf 56}, 2039.\\ 
HERNEGGER, F.  1972  {\it Proceedings of the V European
		 Conference on Controlled Fusion and Plasma Physics},
		 (EURATOM-CEA, Grenoble), V. I, p. 4.\\
KERNER, W. \&   TOKUDA, S. 1987  {\it  Z. Naturforsch.} {\bf 42a},
	       1154}.  \\ 
L\"UST, R. \& SCHL\"UTER, A. 1957 {\it Z. Naturforsch.} {\bf 12a}, 
	      850. \\ 
MASCHKE, E. K.   1973   {\it Plasma Phys.} {\bf 15}, 535. \\
MASCHKE, E. K. \&  PERRIN, H. 1980   {\it Plasma Phys.} {\bf
		22}, 579. \\ 
MONTGOMERY, D. \& SHAN, X. 1994 {\it Comments Plasma Phys. Contolled 
		 Fusion} {\bf 15}, 315.\\ 
MONTGOMERY, D. \&  BATES, J. W. \& LI, S.  1997a
                {\it Phys. Fluids} {\bf 9}, 1188. \\ 
MONTGOMERY, D. \&  BATES, J. W. \& LEWIS, H. R. 1997b
                {\it Phys. Plasmas} {\bf 4}  1080.\\
MOROZOV, A. I. \& SOLOV\'EV, L. S. 1980   {\it Reviews of 
	       Plasma Physics},  edited by  M. A. Leontovich
	       (Consultants Bureau, New York), Vol. 8, p. 1.\\
PFIRSCH, D. \&   SCHL\"UTER, A. 1962  {\it  Der Einflu\ss  der
		elektrischen Leitf\"ahigkeit auf das 
		Gleichgewichtsverhalten von Plasmen niedrigen Drucks in
		Stellaratoren} 
		Report MPI/PA/7/62, Max-Planck-Institut, Munich 
		(unpublished).\\
PLUMPTON, C. \&  FERRARO, V. C. A. 1955 {\it Astrophys. J.}  {\bf 121},  
	       168. \\ 
SCOTT, S. D.    \& BITTER, M. \& HSUAN, H. \& HILL, K. W.\& GOLDSTON, R. J.
                  \&  VON GOELER, S. \&  ZARNSTORFF, M.     1987
		   {\it Proceedings  of the 14st European 
		   Conference on Controlled Fusion and Plasma
		  Physics  (Madrid 1987)}, 
		  (Geneva:  European  Physical Society) vol 11D, p. 65.\\
SEMENZATO, S. \&  GRUBER, R \&   ZEHRFELD, H. P. 1984  
		{\it Comput. Phys. Rep.} {\bf 1}, 389.\\ 
MILLER, R. L. \& WAELBROECK \& HASSAM, A. B. \& WALTZ, R. E.  
               1995 {\it Phys. Plasmas} {\bf 2}, 3676. \\ 
SEN, S. \& RUSBRIDGE, M. G. 1995 {\it Phys. Plasmas} {\bf 2}, 2705.\\
SHAFRANOV, V. D. 1957 {\it Zh. Eksp. Teor. Fiz} {\bf 33},
	      710 [{\it Sov. Phys. JETP} {\bf 6}, 545 (1958)].\\ 
SHAFRANOV, V. D. 1966 {\it Reviews  of Plasma Physics}
		edited  by  M. A. Leontovich
	       (Consultants Bureau, New York), Vol. 2, p.  116.\\ 
SUCKEWER, S. \& EUBANK, H. P. \& GOLDSTON, R. J. \& 
              HINNOV, E. \& SAUTHOFF, N. R. 1979
              {\it Phys. Rev. Lett.} {\bf 43}, 207.\\ 
TAMMEN, H. F. \& DONN\'{E}, A. J. H. \& EURINGER, H. \& OYEVAAR, T.
                1994 {\it Phys. Rev. Lett.} {\bf 72}, 356.\\ 
TASSO, H. 1970 {\it Phys. Fluids} {\bf 13}, 1874. \\                 
TASSO, H.  1979 {\it Lectures on Plasma Physics},
		Report  IFUSP/P-181, LFP-8,  Universidade 
		de S\~ao Paulo, Instituto de F\'isica, S\~ao Paulo.\\
THROUMOULOPOULOS, G. N. \&   PANTIS, G. 1989 
               {\it Phys. Plasmas B}  {\bf 1 }, 1827. \\
THROUMOULOPOULOS, G. N. \&   PANTIS, G. 1996 
	       {\it Plasma Phys. Control. Fusion} {\bf 38}, 1817. \\ 
THROUMOULOPOULOS, G. N. \&   TASSO, H.  1997   
		 {\it Phys. Plasmas} {\bf 4}, 1492.\\ 
WESSON, J. 1987  {\it Tokamaks} (Clarendon Press, 
		 Oxford), p. 88.\\
ZEHRFELD, H. P. \& GREEN, B. J. 1972 {\it Nucl. Fusion} {\bf 12}, 569.\\
\.ZELAZNY, R \& STANKIEWICZ, R. \& GA LKOWSKI, A. \& POTEMPSKI, S.
            1993 {\it Plasma Phys. Contr. Fusion} {\bf 35}, 1215.\\
\newpage
\begin{center}
{\large\bf Figure}
\end{center}
\vspace{15cm}

\noindent
FIG 1 The geometry of an axisymmetric toroidal plasma surrounded 
      by a conductor.
\end{document}